# Analyzing Confidentiality and Privacy Concerns: Insights from Android Issue Logs

Sherlock A. Licorish
Department of Information Science
University of Otago
PO Box 56, Dunedin 9054, New Zealand
sherlock.licorish@otago.ac.nz

Stephen G. MacDonell and Tony Clear
SERL, School of Comp. & Math. Sciences
Auckland University of Technology
Auckland 1142, New Zealand
smacdone@aut.ac.nz, tony.clear@aut.ac.nz

## ABSTRACT

**Context**: Post-release user feedback plays an integral role in improving software quality and informing new features. Given its growing importance, feedback concerning security enhancements is particularly noteworthy. In considering the rapid uptake of Android we have examined the scale and severity of Android security threats as reported by its stakeholders. **Objective**: We systematically mine Android issue logs to derive insights into stakeholder perceptions and experiences in relation to certain Android security issues. **Method**: We employed contextual analysis techniques to study issues raised regarding confidentiality and privacy in the last three major Android releases, considering covariance of stakeholder comments, and the level of consistency in user preferences and priorities. **Results**: Confidentiality and privacy concerns varied in severity, and were most prevalent over Jelly Bean releases. Issues raised in regard to confidentiality related mostly to access, user credentials and permission management, while privacy concerns were mainly expressed about phone locking. Community users also expressed divergent preferences for new security features, ranging from more relaxed to very strict. **Conclusion**: Strategies that support continuous corrective measures for both old and new Android releases would likely maintain stakeholder confidence. An approach that provides users with basic default security settings, but with the power to configure additional security features if desired, would provide the best balance for Android's wide cohort of stakeholders.

## Categories and Subject Descriptors

D.2.9 [**Software Engineering**]: Management.

**General Terms:** Quality, Security, Performance.

**Keywords:** Android, Security, Confidentiality, Privacy, Content Analysis, Empirical Analysis.

## 1. INTRODUCTION

Post-release end-user feedback plays an integral role in improving software quality [1]. Through end-users' feedback developers are able to gauge their sentiments about released products. In some instances users are also able to rate software, which may inform other users' decisions. Furthermore, apart from improving the quality of previously deployed software features, post-release feedback also signals other desired functionality, and so can direct a software product's evolution. Insights from highly successful cases could identify critical success factors for others.

The Android operating system (OS) has arguably become the most widely adopted mobile OS [2]. In recent times, however, there has been growing unease regarding the quality of the Android platform [3]. In particular, security-related concerns have become the focus of user reviews [3]. This is driven, in part, by the increasing capabilities of mobile devices, with users now able to store non-trivial amounts of private data on their mobile handsets, along with the greater use of mobile devices in corporate settings under bring-your-own-device (BYOD) policies.

Given the critical role security plays in assessments of software quality [4], there is need for research to explore how such issues have affected the Android OS. Beyond uncovering post-release insights into a highly successful software product, such explorations could also provide direction to developers in terms of employing suitable precision when developing countermeasures for particular security threats. In addition, insights regarding the scale and severity of various Android security threats could ensure new customer awareness. In order to ascertain the possible utility of such a study we conducted a preliminary analysis of the Android community's concerns and found that 79% of Android users' security-related comments related to either confidentiality or privacy. Understanding the nature of these issues, whether they covary, and any variance in the community's preferences and priorities, could usefully inform remedial efforts.

While the Android OS and its in-built issue tracker have attracted previous research efforts, there has been a tendency towards manifest (surface) level analysis [5, 6]. We believe that, though useful, such efforts only reveal a part of the picture, and so should be supplemented by deeper contextual analyses. Moving beyond analyses based on word use frequency, qualitative forms of contextual analysis enable researchers to assess communicators' *intentions* and the implications of these intentions on a process or construct [7]. Such an approach would therefore help us to unpack the details reported in the Android issue tracker, and provide insights into the abovementioned issues. We therefore used contextual analysis approaches to examine the details of Android issues as logged by stakeholders over the last three major Android OS releases: Ice Cream Sandwich (4.0, 4.03), Jelly Bean (4.1, 4.2, 4.3) and KitKat (4.4). We provide empirical evidence of the nature and scale of confidentiality and privacy issues facing the Android community, how the mix of these issues changed over the three releases, and stakeholders' expressed preferences and competing concerns. Our contributions are threefold, and should support quality improvements: we discuss our findings in relation to previous evidence and technical opinions, we identify strategies for counterbalancing various stakeholders' demands for fixes, and we outline implications for the mobile community.

In the next section we provide our study background and state our research questions, and we then describe our research setting in Section 3. In Section 4 we present our results, and in Section 5 we discuss our findings and outline potential strategies. We then consider the threats to the work in Section 6, before outlining implications for the mobile community and providing concluding remarks in Section 7.

## 2. BACKGROUND AND QUESTIONS

There is clearly a strong imperative for software producers to consider post-release reviews of their products. Previous work considering the acceptance of users' concerns and opinions on such products has long established that software (and any associated hardware) is most successful when end-user feedback is accommodated [8, 9]. Literature examining the relationship between end-users' participation and software product success has also linked the acceptance of end-user feedback to their satisfaction with or acceptance of the product [10]. Willingness to accommodate end-user feedback has also been shown to affect the influence of the delivered system on the end-user community [11].

In delivering and sustaining software product quality, security-related issues may be particularly noteworthy, as they are likely to require urgent action from developers. Previous studies have lent some support to this proposition. For instance, Zaman et al. [4] compared developers' focus on security and performance bugs in Firefox and found that security bugs were favored for fixing over those that were performance-related, and were fixed much more quickly. Critical security bugs have also been removed from the Android issue list to avoid or reduce exacerbation or exploitation of such issues [12].

At the core of the Android OS stack is a modified Linux 2.6 monolithic kernel, with Java applications running on a virtual machine [2]. Among the software programs that are shipped as part of the Android OS, the Contacts application, Email client, Web and Map browsers and Messaging application are those most frequently included in vendor instantiations. Multiple handset vendors collaborate with Google through the Open Handset Alliance (OHA), extending these applications (i.e., how the features appear), and the basic Android OS, to suit their hardware offerings. Companies such as HTC, Samsung, LG and Sony are among the device manufactures that offer Android phones, while Sprint, Verizon, T-Mobile and AT&T offer services for Android devices. These communities, along with other developer groups, regular end-users, and Google itself, use the Android OS issue tracker to report post-release issues and request enhancements to features. Thus, the Android issue tracker provides the interface between the Android OS (as the product), the producers of the product (the Google developer community) and the consumers of the product (device vendors, app and service developers and end-users of Android devices).

Researchers have thus examined this interface to understand various aspects of the Android OS. For instance, Kumar Maji et al. [13] studied issues reported for four early versions of the Android OS (versions 1.1, 1.5, 1.6 and 2.0) and found most defects to be present in the application layer. Guana et al. [5] classified 8,597 Android OS issues in four layers of this OS (application framework, library, android runtime and Linux kernel), omitting those that were suspected to be in the application layer. They found higher levels of defect concentration in the framework and kernel layers. Guana et al. [5] suggested that this prevalence of defects closer to the OS kernel may be linked to hardware compatibility issues.

As noted in Section 1, with security being seen as central to user perceptions of software quality [4], leading in part to growing interest in the security of mobile OSs [12, 14], it would seem timely to explore and provide understandings for the nature of Android security-related issues, how stakeholders' views covary, and various users' preferences and priorities with respect to changes. Such insights would provide indirect understandings around the attention that is given to such issues by the community. Furthermore, with Android devices leading mobile device sales [2], understanding the frequency with which security issues are raised in the current Android OS offerings would likely support users' confidence. We thus examine confidentiality and privacy issues raised on the Android issue tracker in order to answer the following questions:

RQ1. What is the scale of confidentiality and privacy issues raised for the Android OS versions?

RQ2. Are specific versions of Android OS more issue-prone than others?

RQ3. Are stakeholder views regarding confidentiality and privacy issues homogeneous or are they likely to create dilemmas for Android developers?

## 3. RESEARCH SETTING

Issues identified by the Android community are submitted to the Android OS issue tracker hosted by Google; refer to http://code.google.com/p/android/issues/list. Among the data that is stored in the issue tracker are the following details: Issue ID, Type, Status, Owner, Summary description, Stars (number of people following the issue), Priority, Milestone, Attachments, Open date, Close date, Reporter, Reporter Role, Project, Component, and OS Version. We extracted a snapshot of the issue tracker, comprising 21,547 issues logged between January 2008 and March 2014. These issues were then imported into a database, and thereafter, we performed data cleaning by executing previously written scripts to remove all HTML tags and foreign characters [15, 16], and particularly those in the Summary description field, to avoid confounding of our analysis.

We next employed exploratory data analysis (EDA) techniques to investigate the data properties and to facilitate anomaly detection. We observed that issues were labelled as defect (15,750 issues), enhancement (5,354 issues) and others (5 issues); and 438 issues had no type (being null). Issues had one of six statuses: new (18,891 issues), needsinfo (143 issues), unassigned (476 issues), assigned (2,001 issues), resolvedbyuser (1 issue) and accepted (32 issues). Three issues also had the null status. Issues had 140 different owners. They were logged mostly by those identifying themselves as users (9,006 issues) and developers (7,804 issues); with some 4,737 issues being entered anonymously. Issues were reported for 13 different components, although for most of the issues reported this field was left blank (15,711 issues altogether). We observed that only 2,816 issues had the version field updated

(out of the total 21,547 issues), while the others were left blank. Given this, we did not perform extensive analysis on data columns with missing values.

We examined the data of each issue in our database to correlate these with the commercial releases of the Android OS (refer to http://www.android.com). Its first release was in September 2008 (http://android-developers.blogspot.co.nz/2008/09/announcing-android-10-sdk-release-1.html/), while the first issue was logged in the issue tracker in January 2008. This suggests that the community was already actively engaged with the Android OS after the release of the first beta version in November 2007 (refer to http://android-developers.blogspot.be/2007/11/android-first-week.html/), with issues being reported just two months after the first beta release. Given this level of active engagement and issue identification, occurring even before the official Android OS release, we partitioned the issues based on Android OS release date and major name change. So, for instance, all of the issues logged from January 2008 (the date of the first issue that was entered on the issue tracker) to February 2009 (the date of one of the Android releases before a major name change was made) were labelled as 'Early versions', reflecting the period of the Android OS releases 1.0 and 1.1 which were both without formal names. The subsequent partition comprised the period between Android OS version 1.1 and Cupcake (Android version 1.5), and so on.

Table 1 provides a summary of the numbers of issues that were logged between each of the major releases, from the very first commercial release (and using the release date of the first beta version to compute the first entry) to KitKat – Android version 4.4. From column three of Table 1 (Number of days between releases) it is noted that the time taken between the delivery of most of Android OS's major releases (those involving a name change) fell between 80 and 156 days, with three of the ten releases (Early versions, Gingerbread and Jelly Bean) falling outside this range. The fourth column of Table 1 (Total issues logged) shows that the number of issues reported increased somewhat as the Android OS progressed, with this rise being particularly evident when the mean number of issues reported per day for each release is considered (refer to the values in the fifth column for details). Over the six years of Android OS's existence, on average, 9.6 issues were logged every day (median = 4.4, Std Dev = 13.6).

As noted in Sections 1 and 2, we were especially interested in the security (confidentiality and privacy) issues that were reported for the last three releases (as per the highlighted cells in Table 1). Our selection of these three releases is driven by continued consumer demand for these offerings (http://www.cnet.com/news/kitkat-chews-up-more-than-20-percent-of-android-devices/), and by our wish to provide actionable recommendations for the mobile stakeholder community. We discuss the approach used for extracting security-related issues and our analysis methods in the next three subsections.

### 3.1 Classifying Security Issues

Bhattacharya et al. [12] identified 980 bug reports in the Android OS by querying words such as "security", "vulnerability", "attack", "crash", "buffer overflow" and "buffer overrun". Other security terms included under the ISO9126 quality model functionality category, and used by Hindle et al. [6] in their evaluation of MaxDB and MySQL, include "exploit", "certificate", "secured", "malicious" and "trustworthy". The mainstream OS literature generally considers multiple areas of security, including privacy, confidentiality, integrity, availability and reliability [17, 18]. Privacy denotes a state of being free from intrusion; confidentiality relates to limiting unauthorized access. Integrity denotes freedom from corruption; the state of being available is defined as being accessible. Finally, reliability denotes the state of being dependable. We anticipated that a classification scheme considering these five areas would capture a broader spectrum of security issues than had been considered in previous studies (e.g., refer to [12]), and would also provide more granular separation of security issues, although the terms considered under each area still converge with those of the ISO9126 quality model. Informed by these various threads in the literature we thus created the classification scheme covering these five areas (shown in Table 2) to classify Android OS security issues.

We tokenized the Summary description field of the issues into word unigrams and, based on the classification scheme in Table 2, we then extracted all of the security-related issues in our snapshot of the Android issue tracker. We then visualized these results, which revealed that, of the security-related concerns captured by our protocol, those relating to privacy (36.7%) and confidentiality (42.1%) dominated the issues raised, as depicted in Figure 1. We thus scrutinize these two subsets of issues using the following contextual analysis approaches.

### 3.2 Conventional Content Analysis

Our classification scheme in Table 2 identified 510 issues relating to confidentiality and 1103 privacy-related issues in the Android issue tracker over the last three OS releases (Ice Cream Sandwich, Jelly Bean and KitKat). We first selected the smaller sample of 510 confidentiality issues for open coding using conventional content analysis. In this phase of coding we decided to use a bottom-up approach, where codes were derived from the issues as against using a predefined coding scheme. Researchers employing such an approach generally start the process of data analysis by inductively examining the data, allowing meaning to flow from the data, as against approaching data analysis with any preconceptions [19]. Two coders (the first author and another trained coder) initially perused 5% of the confidentiality issues (26 in total), and assigned each to a topic. During this exercise it was observed that, quite frequently, each issue addressed a single topic, and was of four types: 1. Feature does not work as intended; 2. Feature violating constraint; 3. Need for new feature; or 4. Feature does not work (see summary categories, examples and frequencies in Table 3). In addition, we observed that issues categorized as enhancement requests (i.e., 3. Need for new feature) sometimes reflected competing concerns, which would have the potential to create dilemmas for developers in terms of their deciding on appropriate fixes. We then recoded all 1613 issues in a formal coding phase (around 5% had dual concerns), with each issue being assigned to one of the four types just noted (refer to Table 3 for a summary). We then performed formal reliability assessment, which revealed that there was 88% inter-rater agreement between the two coders as measured using Holsti's coefficient of reliability measurement (C.R) [20]. The remaining coding differences were discussed and resolved by consensus. Our reliability measure represents excellent agreement between coders and suggests that a consistent and reliable approach was being taken. The enhancement requests (those issues that were coded as Scale 3 in Table 3) were then probed further using the analysis approach outlined next (refer to Section 3.3). Prior to conducting this additional round of analysis on the enhancement requests, however, we undertook a number of landscape analyses to extract meaning from the issues raised, the results of which are provided in Sections 4.1.1 and 4.2.1.

## 3.3 Dilemma Analysis

As noted above, during our content analysis confidentiality and privacy issues that suggested new features and directions for improving the Android OS were coded as enhancements. However, informal perusal of these enhancements also revealed that there were some conflicting requests from the Android stakeholders. Our first round of content analysis did not capture these interpretations fully, but was instead closer to the surface of the issues. We anticipated, however, that a deeper examination of these conflicting requests would reveal competing concerns in the Android community. Dilemma analysis, often referred to as the sociological conception of contradiction, can be used to unpack opposing points of view [21]. This approach guides the analysis of transcripts to extract *issues* about which individuals hold opinions. In our context, while the issues were not recorded as transcripts as such, the enhancement requests contained sufficient detail to enable us to both identify the new feature requested and the potential benefit of having such a feature. Thus, it was straightforward for us to identify competing concerns among such issues. We thus examined each enhancement request, paying close attention to those that conflicted with other issues. These results are provided in Sections 4.1.2 and 4.2.2.

## 4. RESULTS

We separate the results for the two sets of security issues in Sections 4.1 and 4.2. We first present our findings for the confidentiality-related issues in Section 4.1. We then provide our findings for those related to privacy in Section 4.2.

## 4.1 Confidentiality-Related Issues

We first outline the results from our conventional content analysis in Section 4.1.1. We then examine the competing concerns in Android stakeholders' confidentiality-related enhancement requests, and provide these results in Section 4.1.2.

### 4.1.1 Content Analysis: Confidentiality

Of the 510 confidentiality-related issues raised on the Android issue tracker, stakeholders identifying themselves as users registered 224 complaints, those registered as developers lodged 187 concerns and another 99 issues were recorded anonymously.

The largest number of confidentiality issues were labelled as defects (388), and 122 issues were logged as enhancements. Figure 2 (a) shows how these issues were distributed by confidentiality terms overall (refer to Table 2 for details), where it is revealed that issues related to *access*, *username, password* and *permission* dominated these concerns in the Android issue tracker over the latter three major releases (Ice Cream Sandwich, Jelly Bean and KitKat). A Pearson Chi-square test was conducted to ascertain whether the differences observed in Figure 2 (a) were statistically significant. The results of the Chi-square test confirm that there were significant differences in the types of issues that were recorded on the Android issue tracker, and particularly for the higher levels of *access*-, *password*-, and *permission*-related issues that were logged ($X^2 = 68.08$, df = 36, $p < 0.01$).

Given this finding, we considered how those issues were distributed across the three releases concerned, and depict the results in Figure 2 (b). Here it is shown that most issues were raised over the course of the Jelly Bean releases, with concerns about *access*, *login/username*, *password*, *permission* and *verification* dominating those issues recorded. We also observe in Figure 2 (b) that there has been heightened concern about restriction since the last Android release (KitKat), and that between the periods of the release of the Ice Cream Sandwich and Jelly Bean versions stakeholders recorded the fewest confidentiality-related issues. We again performed a Pearson Chi-square test to ascertain whether the differences observed in Figure 2 (b) were statistically significant, considering the seven most popular issues in Figure 2 (b) in our test. Our Chi-square test confirms that there were significant differences in the number of and types of confidentiality issues recorded over the latter three major versions of the Android OS ($X^2 = 25.44$, df = 6, $p < 0.01$).

We present a finer grained analysis of these results in Table 4, considering the frequency (including percentages – though we caution on the interpretation of percentages given the low frequency of some issues) of issues for the most regularly reported *access*, *password* and *permission* confidentiality concerns. Notwithstanding the differences in frequencies, Table 4 shows that for the *access* category, there was substantial variation in the mean number of issues raised over the Jelly Bean releases when compared to those noted after the release of Ice Cream Sandwich and KitKat (67.9% and 74.1% compared to 9.4% and 7.4% for Ice Cream Sandwich and 22.6% and 18.5% for KitKat respectively).

**Table 1. Android OS issues over the major releases**

| Version (Release) | Last release date | Number of days between releases | Total issues logged | Mean issues per day |
|---|---|---|---|---|
| Early versions (1.0, 1.1) | 09/02/2009 | 451 | 262* | 0.6 |
| Cupcake (1.5) | 30/04/2009 | 80 | 101 | 1.3 |
| Donut (1.6) | 15/09/2009 | 138 | 266 | 1.9 |
| Éclair (2.0, 2.01, 2.1) | 12/01/2010 | 119 | 464 | 3.9 |
| Froyo (2.2) | 20/05/2010 | 128 | 490 | 3.8 |
| Gingerbread (2.3, 2.37) | 09/02/2011 | 265 | 1,291 | 4.9 |
| Honeycomb (3.0, 3.1, 3.2) | 15/07/2011 | 156 | 897 | 5.8 |
| Ice Cream Sandwich (4.0, 4.03) | 16/12/2011 | 154 | 1,127 | 7.3 |
| Jelly Bean (4.1, 4.2, 4.3) | 24/07/2013 | 586 | 12,148 | 20.7 |
| KitKat (4.4) | 31/10/2013 | 99 | 4,501 | 45.5 |
| | | $\sum = 2,176$ | $\sum = 21,547$ | $\bar{x} = 9.6$ |

* Total number of issues logged between the first beta release on 16/11/2007 and Android version 1.1 released on 09/02/2009

**Table 2. Security labels and related terms**

| Label | Related terms |
|---|---|
| Privacy | authorization, phone lock, authentication, privacy, seclusion, separateness, isolation, conceal, secure, exploit, prevent, unauthorized, intrusion |
| Confidentiality | secret, classified, privy, permission, password, confidential, vulnerable, access, grant, restrict, verify, privilege, username, login |
| Integrity | corrupt, disrepute, cohesion, coherence, soundness, wholeness, completeness, honest, license, integrity, attack, malicious, modification, identity, detection, sensitivity |
| Availability | accessible, convenient, buffer overflow, buffer overrun, crash, loss, destruction, available, obtain |
| Reliability | trust, reliable, dependable, stable, safe(ty), consistent, certification, validation, performance |

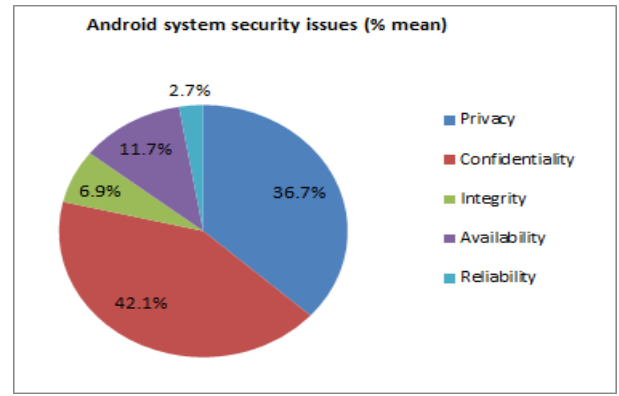

**Figure 1. Android OS security issues**

**Table 3. Coding categories and number of codes**

| Scale | Category/Characteristic | Example | Confidentiality Codes (%) | Privacy Codes (%) |
|---|---|---|---|---|
| 1 | Does not work as intended | "Access to an account with two factor authentication persists after deleting app-specific password used to attach to device" | 146 (28.6) | 440 (39.9) |
| 2 | Violating constraint | "Accessing my call logs takes cool 4-5 seconds" | 132 (25.9) | 175 (15.9) |
| 3 | Need for new feature | "When installing an app: the security permission request to access contact details should be in red or orange and first in the list" | 130 (25.5) | 285 (25.8) |
| 4 | Does not work | "Security exception when accessing account manager data from other apps signed with same keystore" | 102 (20) | 203 (18.4) |

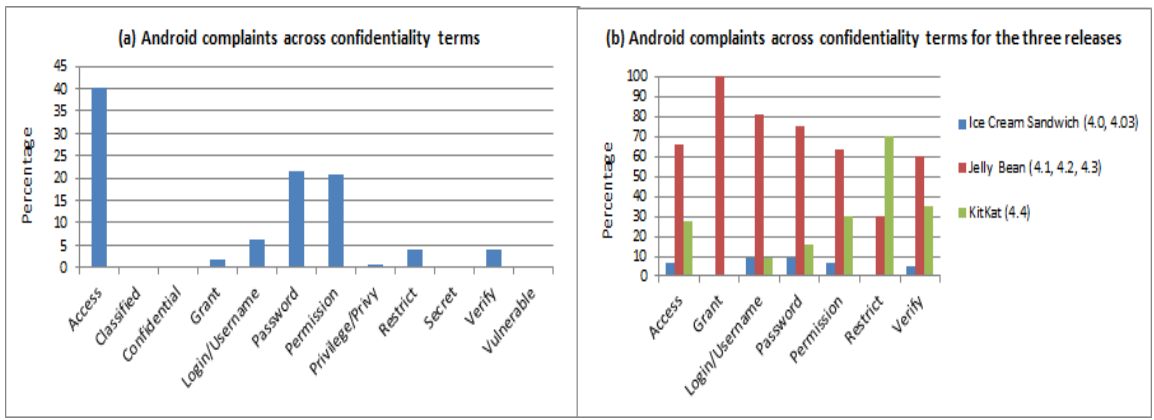

**Figure 2. Android confidentiality issues**

While the logging of *access*-related complaints by stakeholders of Ice Cream Sandwich was below average for all four categories of codes recorded (Scales 1 to 4), since the release of KitKat users have logged more such issues (Scale 1 = 22.6%, Scale 2 = 25.6%, Scale 3 = 40.7% and Scale 4 = 18.5%). We observe in Table 4 that 40.7% (or 24) of the requests for *access*-related features were recorded since the release of KitKat. While there were fewer *password*-related issues raised (as also seen in Figure 2 (a)), Scales 1 and 4 categories of codes were also highest after Jelly Bean releases (85.7% and 77.8% of the codes respectively). This trend of higher numbers of issues raised was replicated for *permission*-related issues. Most of the complaints coded Scale 3 were submitted anonymously (100% for Ice Cream Sandwich, 60% for Jelly Bean and 81.8% for KitKat), while users and developers recorded a similar pattern of codes across versions.

We next take a detailed view of the confidentiality issues that were labelled as enhancement requests (Need for new feature), to assess the level of competing concerns faced by Google developers in delivering on stakeholders' requests.

### 4.1.2 Dilemma Analysis: Confidentiality

We examined the 122 confidentiality enhancement requests to assess the level of stakeholders' competing concerns. Of these, 57 related to access, 6 related to username/login, 26 related to password, 25 related to permission, 4 related to restriction, 1 related to secret and 3 to verification. Ninety-six issues were logged by anonymous users, 16 by developers and 9 by users. Of the confidentiality-related enhancement requests 12 were lodged for Ice cream Sandwich, 67 for Jelly Bean and 43 for KitKat, somewhat replicating the general pattern noted above. Given this small sample of enhancement requests we provide overall results, as against separating the data across versions. These results are summarized in Table 5.

Table 5 reveals that four types of confidentiality concerns (regarding *access*, *login*, *password* and *permission*) demonstrated some form of divergence among stakeholders' preferences, whereas issues related to *restriction*, *secret* and *verification* were homogeneous. In Table 5 it is shown that while some stakeholders were more cautious about how *access* to their data is managed, others were less worried. In fact, one user sought authorized access to stored contact details for some popular apps. On the other hand, another group of users was extremely cautious about any *access* being given to their contacts. Furthermore, while some users were happy to authorize use of their cellular data when there was no wifi connection, another group of stakeholders was seeking more granular control, for example, to be able to grant selected apps permission to use internet data but to restrict others.

Table 5 shows that under the *login* confidentiality keyword there were two issues that saw major divergence: "automatic hotspot login" and "phone restore after wiping". While some stakeholders desired the feature to login to hotspots automatically, others were against this feature, instead opting to trust only some private IPs. Such a split was also evident for the feature to login and restore handsets after wiping. This divergence also extended to the use of *passwords*. While some users favored caching passwords, removing passwords for some VPNs and making passwords visible, others were predisposed to password-protecting the use of wifi, mobile purchases, and even the phone shutdown (see examples in Table 5). For *permission*, there was greater leaning towards granular permission management. Although some users were less strict (e.g., "requesting the need for download without notification"), others felt that more granular permission would increase user confidence.

## 4.2 Privacy-Related Issues

We present the results from our analysis of the privacy-related issues in this section. First, the results from our conventional content analysis are provided in Section 4.2.1. We then examine the competing concerns in Android stakeholders' privacy-related enhancement requests, and provide these results in Section 4.2.2.

### 4.2.1 Content Analysis: Privacy

Of the 1103 privacy-related issues recorded in our snapshot of the Android issue tracker, stakeholders identifying themselves as users registered 647 of these, developers lodged 221 issues and a further 235 were recorded anonymously. The largest number of privacy-related issues were labelled as defects (830 issues), while 273 were logged as enhancements. Figure 3 (a) shows how these issues were distributed by terms that were classified under the privacy category, where it is revealed that issues related to *authentication, lock* and *secure* dominated the Android issue tracker over the latter three major releases. A Pearson Chi-square test was conducted to ascertain whether the differences observed in Figure 3 (a) were statistically significant. We first removed all the entries for terms that had a sample size of less than ten respective codes (the assumption for utilizing a Chi-square test) [22], before executing the test, which confirmed that there were significant differences in the types of privacy issues that were recorded on the Android issue tracker, and particularly for the higher numbers of lock-related issues that were lodged ($X^2 = 28.08$, $df = 9$, $p < 0.01$).

**Table 4. Most regularly reported *access*, *password* and *permission* confidentiality concerns**

| Version | Access(%) | | | | Password(%) | | | | Permission(%) | | | |
|---|---|---|---|---|---|---|---|---|---|---|---|---|
| | 1 | 2 | 3 | 4 | 1 | 2 | 3 | 4 | 1 | 2 | 3 | 4 |
| ICS | 5(9.4) | 3(7.7) | 1(1.7) | 4(7.4) | 2(4.8) | 1(4.8) | 6(21.4) | 1(5.6) | 2(9.1) | 2(4.2) | 3(12.0) | 0(0) |
| JB | 36(67.9) | 26(66.7) | 34(57.6) | 40(74.1) | 36(85.7) | 18(85.7) | 14(50.0) | 14(77.8) | 18(81.8) | 26(54.2) | 17(68.0) | 6(54.5) |
| KK | 12(22.6) | 10(25.6) | 24(40.7) | 10(18.5) | 4(9.5) | 2(9.5) | 8(28.6) | 3(16.7) | 2(9.1) | 20(41.7) | 5(20.0) | 5(45.5) |

ICS=Ice Cream Sandwich, JB=Jelly Bean, KK=KitKat, Scale 1=Does not work as intended, Scale 2=Violating constraint, Scale 3=Need for new feature, Scale 4= Does not work

**Table 5. Confidentiality-related competing concerns**

| Terms | Competing concerns (Grant versus Restrict) | Examples |
|---|---|---|
| Access | Access to other peripheral via USB, Access to contact details, Shared app access to microphone, Apps access to cellular data if wifi not available, Access to SD Card, Network access, Internet access | "Provide anonymized/hashed access to contact details/contacts for instant messagers like e.g. whatsapp, threema, etc." + "Enable cellular data when connected wifi access point doesn't provide internet connectivity" <<>> "When installing an app: the security permission request to access contact details should be in red or orange and first in the list." + "When roaming have an option to decide which apps can get internet access." |
| Login | Automatic hotspot login, Phone (access) restore after wiping | "After fresh login of account (after wiping phone) there no option to configure auto download of apps by device / defer downloads" <<>> "Wipe after consecutive failed login attempts? verify human. enter android to continue" |
| Password | Cache password, Modify encryption pin, Remove VPN password for some wifi, Make password visible, Pin and password optional for VPN, Password protect wifi, Purchase via password protection, Shutdown with password | "Allow for a simpler unlock password/pin than the password/passphrase used for full disk encryption" + "Enable slide unlock until timeout for pin, pattern, or password lock" <<>> "Different passwords for encryption and screen lock" + "User profiles - increase security by password protecting and segmenting" + "Password for purchases" + "provide a way to password-protect shutdown" |
| Permission | Google analytics without permission, Granular permission, Restrict app to local data, App accessing contacts - provide warning, Selected permission when installing apps, There is need for more granular permission, Reverse previously granted permission | "Add download_without_notification to uses-permission drop-down" + "Use of google analytics without asking for internet permission." <<>> "Some permissions are scary and for features people might never use. here's an idea on how to let users who wouldn't use some features still install and use an app." + "Divide read_phone_state permission in two to provide more secure android for users" + "Permission request - reverse list" + "Ability to deny select permission upon app install" |

We next considered how those issues were distributed across the three releases of interest, depicted in Figure 3 (b). We plot the most prominent *authentication*, *lock*, *privacy*, *secure* and *separate/seclude* issues in Figure 3 (b), which reveals that most were logged over Jelly Bean releases, with issues for all of the five keyword categories just mentioned being dominated over this release. This pattern of results is similar to those that were revealed for confidentiality-related issues, where most were logged after Jelly Bean releases. We performed a second Pearson Chi-square test to ascertain whether the differences observed in Figure 3 (b) for the three Android releases were statistically significant, and particularly for the most prevalent *lock* and *secure* issues (comprising 72.4% and 10.3% of the issues overall). Our Chi-square test confirmed that there were significant differences in the number of *lock* and *secure* issues raised across the latter three versions of the Android OS ($X^2 = 17.23$, df = 3, $p < 0.01$), Jelly Bean being the most problematic. Of note also is that the role (user, developer or anonymous) of those logging issues did not affect the pattern of results noted across versions.

We take a more fine-grained look at the measures for *lock* and *secure* issues in Table 6, considering the frequency and scale (including percentages) of these concerns. We are particularly interested in features that did not work as intended (coded Scale 1) or those that did not work altogether (coded Scale 4). Table 6 shows that for the *lock* category, there were substantially more issues logged over Jelly Bean versions when compared to those recorded over Ice Cream Sandwich and KitKat (80.7% and 81.8% compared to 3.1% and 3.6% for Ice Cream Sandwich and 16.2% and 14.6% for KitKat respectively). Additionally, although of a smaller magnitude, *secure* issues were also most prevalent for Jelly Bean. On average, however, more *secure* issues were logged for Ice Cream Sandwich and KitKat (refer to Table 6).

We next provide a detailed view of the privacy-related issues that were labelled as enhancement requests, to assess the level of competing privacy-related concerns.

### 4.2.2 Dilemma Analysis: Privacy

As in 4.1.2 above, we examined the 285 privacy-related issues that were logged as enhancement requests to determine Android stakeholders' competing concerns. Of the set of privacy issues, 10 related to *authentication*, 2 related to *authorization*, 187 were *lock*-related, 6 related to the *prevent* keyword, 13 had the *privacy* keyword, 7 related to *restriction*, 37 were associated with *security* and 23 with *separation*. Users logged 59 of these issues, developers logged 18, and 208 were entered anonymously. In terms of the distribution of enhancement requests across versions, 19 were logged for Ice Cream Sandwich, 208 for Jelly Bean and 58 for KitKat. We examined these issues for competing concerns, and observed that of the eight types of issues, only *authentication*-, *lock*- and *privacy*-related enhancement requests had conflicting preferences. These are summarized in Table 7, which shows that for *authentication*, while some users were keen on enhancing Android's authentication process, favoring a two stage approach, another set of users was in favor of simpler proxy-based authentication. Similarly, under the *lock* category there was divergence in relation to lock mode, the level of locking, access to phone resources when the phone was locked, the storage of user security credentials, the rigor of Android's locking process, the number and enforcement of locking mechanisms, locking of data, and locking of the security menu. For these issues, while some users favored more liberal policies, others were encouraging stricter phone resource monitoring and locking. For instance, one user requested that the handset should be "locked without SIM only at the startup", so that if a SIM is removed after the handset is fully started users should still have access to all phone resources, whereas another user went as far as requesting that Android devices should "randomly shuffle the keys in lockscreen pin screen" and "improve phone lock security establishing maximum number of intents". Under the *privacy* keyword Table 7 shows that there was less divergence, except for how unknown and private numbers are handled. We discuss these findings along with those presented above in the next section.

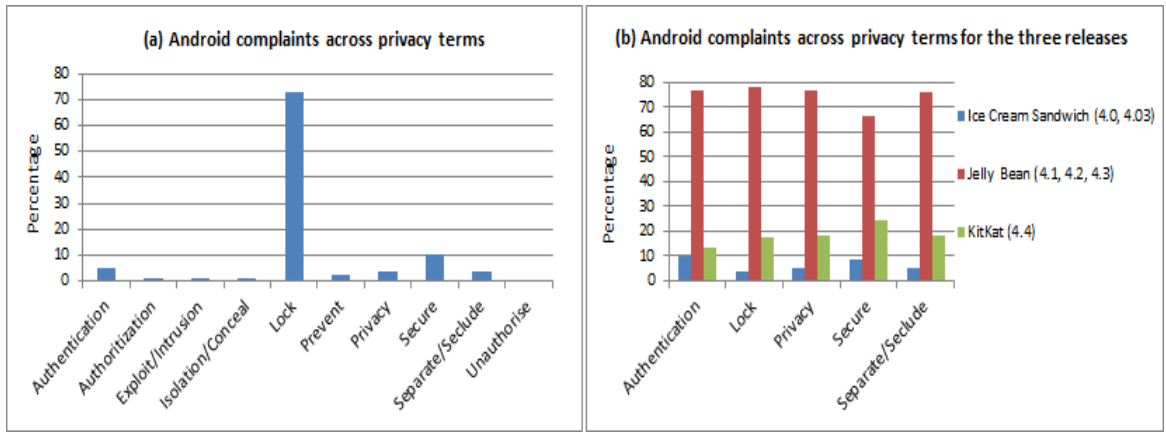

**Figure 3. Android privacy issues**

**Table 6. Most regularly reported *lock* and *secure* privacy concerns**

| Version | Lock(%) | | | | Secure(%) | | | |
| --- | --- | --- | --- | --- | --- | --- | --- | --- |
| | 1 | 2 | 3 | 4 | 1 | 2 | 3 | 4 |
| ICS | 11(3.1) | 5(4.0) | 10(5.3) | 5(3.6) | 0(0) | 2(9.5) | 4(10.8) | 4(13.8) |
| JB | 284(80.7) | 91(70.2) | 143(76.5) | 112(81.8) | 20(74.1) | 15(71.4) | 23(62.2) | 18(62.1) |
| KK | 57(16.2) | 30(23.8) | 34(18.2) | 20(14.6) | 7(25.9) | 4(19.0) | 10(27.0) | 7(24.1) |

ICS=Ice Cream Sandwich, JB=Jelly Bean, KK=KitKat, Scale 1=Does not work as intended, Scale 2=Violating constraint, Scale 3=Need for new feature, Scale 4= Does not work

**Table 7. Privacy-related competing concerns**

| Terms | Competing concerns (Grant versus Restrict) | Examples |
|---|---|---|
| Authentication | Two factor authentication, System wide proxy authentication | "Add a web sign in button to initial set up for two factor authentication users" <<>> "Android system wide proxy authentication" + "Enable proxy authentication for apps" |
| Lock | Lock mode, Start-up lock only if SIM absent, Face unlock, Restricted phone access when phone is locked, Make SIM pin accessible on device and reduce rigour of unlock process, Relax security for specific periods, Multi-lock, Number of intents, Lock screen timer, Data use, Security menu | "Ability to switch between pattern lock and classic lockscreen" + "Make pin or password-locked lockscreen optional when vpn is configured" <<>> "Two level unlock options" + "Unlock by voice" + "Face unlock and slide unlock and vpn" |
| | | "Lock cellphone without sim only at the startup" <<>> "Improve phone lock security establishing maximum number of intents" + "Lock device after a number of fails when trying screen unlocking" + "Randomly shuffle the keys in lockscreen pin screen" |
| | | "Face unlock: secondary picture" + "Ability to retain or save facial training for face unlock if switching to different unlock method" + "Option to skip face unlock" + "Face unlock user auto detect" <<>> "Implement fingerprint reading (via the touch screen) as a more secure alternative to face unlock" + "Face unlock: scan user's retinas with front camera as an added layer of security" |
| | | "Ability to view read-only notifications on lock screen when protected with a pattern/passcode" + "Add camera to locked screen" + "Open custom applications from lock screen" + "Integrating music control on lockscreen" <<>> "Option to disable going to camera from lockscreen" + "Need a lock screen without widgets and without any possible actions on the status bar" + "Option to remove "power off" and "airplane mode" from lockscreen to enhance smartphone security." |
| | | "Enable slide unlock until timeout for pin, pattern, or password lock" <<>> "Improve encrypted device behavior: after 24h w/o unlocking assume stolen and do encrypted suspend to disk or power-off" + "Option to wipe device after a number of incorrect unlocking attempts" + "Lock screen timer timeout to start after inactivity" + "Automatic data switch off on screen lock" + "Add unlock code to access in security menu" |
| Privacy | Unknown/private number | "Allow blocking sms from unknown/private numbers" <<>> "Sms: enable private numbers and numbers as normal characters" |

## 5. DISCUSSION

*RQ1. What is the scale of confidentiality and privacy issues raised for the Android OS versions?* The 1613 issues in our snapshot that were found to be related to confidentiality and privacy amount to less than one percent of the 17,776 issues that were recorded across the latter three Android releases, perhaps suggesting that such issues were infrequent, and therefore unimportant. However, given the critical nature of security-related concerns, the existence of *any* such issues could still negatively impact quality perceptions. This position is particularly supported by previous evidence that has noted that fixes for security issues take longer than those that were otherwise classified [12]. Of the four categories of codes that emerged from our content analysis process, we observed that the highest number of stakeholders' issues was recorded to the "does not work as intended" category. Similarly, over 25% of the confidentiality and privacy issues logged on the Android issue tracker outlined stakeholders' desires for new security features. We also note that there were more than twice as many privacy issues as confidentiality issues raised. This finding is revealing considering that Android users were previously held to be minimally aware of such issues [23]. Overall, we observed an increase in the number of stakeholders' issues raised in the latter releases of the Android OS. We anticipate that this pattern may be linked to increasing capability and complexity of Android devices and their associated OSs, as well as to Android's growing market share. For instance, the early T-Mobile G1 (Android 1.0) device possessed basic hardware and software capability, and had no on-screen keyboard or multi-touch capability, whereas the recent Nexus 5 (Android 4.4) provides these capabilities, along with advanced resource management and optimization (for CPU, memory and I/O), multi-mode processing, enhanced security and across-the-board application integration (e.g., Contacts, Gmail and SMS). However, this finding might also have been tempered, as latter Android releases were generally held to be more security-focused, including using the SELinux access control system (http://tinyurl.com/pyvb3he).

*RQ2. Are specific versions of Android OS more issue-prone than others?* Our results show that some versions of the Android OS led to more issues being raised than others. From a confidentiality perspective, Android stakeholders were most concerned about *access*, their *credentials* and the management of *permission* to their phone resources over the three major releases considered (Ice Cream Sandwich, Jelly Bean and KitKat). In regard to privacy, stakeholders logged most issues about *authentication*, phone *lock* and their phone resources being *secure*, with phone *lock* issues being especially pronounced. Stakeholders recorded the most confidentiality and privacy issues over the Jelly Bean releases. The *lock*-related issues for "does not work as intended" and "does not work" were particularly dominant over Jelly Bean releases compared to Ice Cream Sandwich and KitKat. This pattern of higher prevalence of issues in Jelly Bean may be related to its higher level of usage [2, 24], and Android's quest to deliver superior mobile capability to that offered by its competitors may also have impacted Google's aggressive release cycles. Strikingly, however, there were many major bug fixes delivered as part of the Jelly Bean releases (http://tinyurl.com/pv79q5d). These fixes were probably influenced by the high level of end-user complaints, as seen in our results in terms of the number of issues that were reported over these versions. In addition, KitKat is installed on a larger cohort of Android devices than Jelly Bean

(http://tinyurl.com/palhx7q/), which suggests that the latter versions of the OS were indeed potentially more problematic.

*RQ3. Are stakeholder views regarding confidentiality and privacy issues homogeneous or are they likely to create dilemmas for Android developers?* Android stakeholders were not homogeneous in terms of their desire for confidentiality-related enhancements around *access*, *login*, *password* and *permission*. Rather, while some users were cautious about how access to their data is managed, others were less worried. This divergence could be problematic for those responsible for strategically directing Android's offerings. In addition, this spread of preference also points to variation in end-users' orientation and to varying levels of security awareness in the Android community [3, 25]. In fact, while some stakeholders' requests are likely to create a problem for other users if these were implemented by Google (e.g., some users requested a feature to manage which apps are able to use the internet; however, the need to actually manage such a granular level of security could be annoying to others), others could create or heighten a device's vulnerabilities (e.g., "automatic hotspot login"). We also observe variations in users' willingness to be systematic, which could also create burdens for the Android community. For instance, while some stakeholders were happy to quickly access their device in its previous usable state should it be wiped, and so, wanted to download previously installed apps once acquiring the recently erased handset, others were more cautious, opting for a phased and controlled phone restore. This divergence also extended to the use of passwords. There is likelihood that a previously installed app(s) could have been the source of the security breaches that resulted in the phone wipe in the first instance, and thus, a hasty reinstallation could be ill-advised.

We also observed some conflicting privacy-related requests, particularly those related to *authentication*, *lock* and *privacy*. While some users requested enhancements to Android's authentication process, favoring a two stage approach, another set of users was in favor of simpler proxy-based authentication, potentially creating similar dilemmas to those mentioned above. In fact, under the *lock* category there was divergence in terms of lock mode, the level of locking, access to phone resources when the phone is locked, the storage of user security credentials, the rigor of Android's locking process, the number and enforcement of locking mechanisms, locking of data, and locking of the security menu. In terms of competing concerns for the management of permission, there was greater leaning towards granular permission management. Although some users were less strict, others felt that more granular permission management would increase user confidence. Others were also promoting the idea of reversing previously granted permissions and overriding some default requested permissions during app installation.

The management of permissions has been shown to challenge most mobile users [3], and so the demand for additional management control seems impractical. However, perhaps the power to override previously granted permissions may be helpful to (some) stakeholders. Such a move would assume awareness of resource violations, however, and it has been shown that on many occasions users are unaware of malicious software exploiting their resources [25]. One alternative would be for a centralized audit to be performed by Google from time to time, to assess phone activity logs for malicious activity. Through such an audit malicious software may then be flagged or removed. While this will require internet (data) usage and remote access to devices, users may trade-off such issues with the increased security that would result. Of course recent efforts by Google to add a face unlock feature (in Ice Cream Sandwich), data usage analysis monitors (in Jelly Bean) and modular update (in Jelly Bean), would also reduce security threats, thereby improving Android's quality. We examine these issues further in Section 7.

# 6. THREATS TO VALIDITY
While we have examined an important topic area, and have provided insights into Android OS's confidentiality and privacy issues, there are shortcomings to this work that may affect its generalizability. We consider these in turn.

Although the Android issue tracker is publicly hosted, and so is likely to capture most of the community's concerns [13], issues may also be informally communicated to and addressed within the development teams at Google. Similarly, unreported issues are not captured by our analysis. We also focused as far as possible to include all terms and their synonyms to examine the concepts that were under consideration [6]. However, we accept that there is a possibility that we could have missed some relevant terms. That said, the convergence of our results (revealed through multiple contextual analysis techniques) triangulated our classification scheme, and suggests that our approach was generally robust. In fact, our reliability assessment measure revealed excellent agreement between coders, suggesting that our findings benefitted from accuracy, precision and objectivity [20].

We separated the issues based on the dates of the major Android OS releases. Given that device manufacturers have been shown to delay upgrading their hardware with recent Android OS releases [26], there is a possibility that some issues reported between specific releases were in fact related to earlier releases. However, this misalignment was not detected during our contextual analysis, suggesting that our approach appropriately classified issues.

Finally, although the issue trackers of many mobile OSs are not publicly available, and the distribution of these OSs' issues may not be similar to what is observed in this work for the Android OS, mobile OSs such as Microsoft Windows, Apple iOS, Symbian and BlackBerry are all likely to follow release-maintenance cycles similar to that of Android OS in order to remain competitive in the market.

# 7. IMPLICATIONS AND CONCLUSION
Our findings in this work have implications for Android community stakeholders. For instance, with Google's release of the Nexus 4 for KitKat (and Nexus 5, 6 and 9 for Android Lollipop released on June 25, 2014), developers are likely to give priority to fixes on these OS versions given the need to quickly address stakeholder concerns on the new platforms. Thus, older devices that continue to be shipped with Ice Cream Sandwich and Jelly Bean are likely to inherit any reported security vulnerabilities if these are not explicitly addressed. While there were few reported threats for Ice Cream Sandwich releases, the opposite was seen after the releases of Jelly Bean. Thus, Android end-users should take this into consideration when acquiring new devices. Perhaps a valid strategy for remedial work by Google developers should be to prioritize issues regarding features that "do not work as intended" and those that "do not work". In fact, although small relative to the number of issues that were reported for Android overall, a strategy that focuses on addressing security issues should help to instill stakeholders' confidence in the quality of the Android product range. Our evidence suggests that phone lock and access, user credentials and the management of handsets' permissions would be useful areas for consideration in Google's maintenance strategy. Stakeholders are also expressing growing concern about phone restrictions for the recent KitKat release, and so this issue should similarly be given priority.

Our evidence suggests that, in terms of expanding Android OS security features, Google may face dilemmas in deciding to whose views they should assign most weight. We observed that while some users were cautious about how access to their device resources is managed, others were less worried, in fact requesting relaxed security procedures. This divergence in preferences presents competing concerns among users, and so would need to be carefully managed to satisfy Android's diverse user cohort. Perhaps a strategy to provide users with basic default security settings, with the power to configure additional security features if needed or desired, would provide the best balance. Such an arrangement would allow those who are more security-conscious to enact rigorous controls to protect their privacy, while others who are less concerned may accept minimum security settings. For example, for the "phone restore after wiping" issue, while some stakeholders were happy to quickly access their device in its previous usable state, others were more cautious, opting for a phased and controlled phone restore. Thus, a routine that allows users to configure either of these options would satisfy both groups. In the same way, a strategy that enables users to configure whether or not to enforce "password protection of mobile purchases and mobile device shutdown" would satisfy both groups' desires. Furthermore, such strategies could be supplemented by the power to override previously granted permissions along with centralized audits performed by Google for malicious activity in order to remove such threats.